\documentclass[12pt]{article}
\catcode`\@=11

\@addtoreset{equation}{section}
\def\theequation{\thesection\arabic{equation}}

\def\@normalsize{\@setsize\normalsize{15pt}\xiipt\@xiipt
\abovedisplayskip 14pt plus3pt minus3pt%
\belowdisplayskip \abovedisplayskip
\abovedisplayshortskip  \z@ plus3pt%
\belowdisplayshortskip  7pt plus3.5pt minus0pt}
\def\small{\@setsize\small{13.6pt}\xipt\@xipt
\abovedisplayskip 13pt plus3pt minus3pt%
\belowdisplayskip \abovedisplayskip
\abovedisplayshortskip  \z@ plus3pt%
\belowdisplayshortskip  7pt plus3.5pt minus0pt
\def\@listi{\parsep 4.5pt plus 2pt minus 1pt
            \itemsep \parsep
            \topsep 9pt plus 3pt minus 3pt}}

\def\underline#1{\relax\ifmmode\@@underline#1\else
        $\@@underline{\hbox{#1}}$\relax\fi}
\@twosidetrue \relax

\catcode`@=12

\catcode`\@=11

\def\section{\@startsection{section}{1}{\z@}{3.5ex plus 1ex minus
   .2ex}{2.3ex plus .2ex}{\large\bf}}
\def\thesection{\arabic{section}.}

\def\ps@headings{\def\@oddfoot{}\def\@evenfoot{}
\def\@oddhead{\hbox{}\hfill
        \makebox[.5\textwidth]{\raggedright\ignorespaces --\thepage{}--
        \hfill }}
\def\@evenhead{\@oddhead}
\def\subsectionmark##1{\markboth{##1}{}}
}

\ps@headings

\catcode`\@=12

\relax

\def\figcap{\section*{Figure Captions\markboth
        {FIGURECAPTIONS}{FIGURECAPTIONS}}\list
        {Fig. \arabic{enumi}:\hfill}{\settowidth\labelwidth{Fig. 999:}
        \leftmargin\labelwidth
        \advance\leftmargin\labelsep\usecounter{enumi}}}
 \relax
\def\tablecap{\section*{Table Captions\markboth
        {TABLECAPTIONS}{TABLECAPTIONS}}\list
        {Table \arabic{enumi}:\hfill}{\settowidth\labelwidth{Table 999:}
        \leftmargin\labelwidth
        \advance\leftmargin\labelsep\usecounter{enumi}}}
 \relax
\def\reflist{\section*{References\markboth
        {REFLIST}{REFLIST}}\list
        {[\arabic{enumi}]\hfill}{\settowidth\labelwidth{[999]}
        \leftmargin\labelwidth
        \advance\leftmargin\labelsep\usecounter{enumi}}}
 \relax

\catcode`\@=11

\def\marginnote#1{}
\newcount\hour
\newcount\minute
\newtoks\amorpm
\hour=\time\divide\hour by60 \minute=\time{\multiply\hour by60
\global\advance\minute by- \hour}
\edef\standardtime{{\ifnum\hour<12 \global\amorpm={am}%
    \else\global\amorpm={pm}\advance\hour by-12 \fi
    \ifnum\hour=0 \hour=12 \fi
    \number\hour:\ifnum\minute<100\fi\number\minute\the\amorpm}}
\edef\militarytime{\number\hour:\ifnum\minute<100\fi\number\minute}
\def\draftlabel#1{{\@bsphack\if@filesw {\let\thepage\relax
  \xdef\@gtempa{\write\@auxout{\string
    \newlabel{#1}{{\@currentlabel}{\thepage}}}}}\@gtempa
    \if@nobreak \ifvmode\nobreak\fi\fi\fi\@esphack}
     \gdef\@eqnlabel{#1}}
\def\@eqnlabel{}
\def\@vacuum{}
\def\draftmarginnote#1{\marginpar{\raggedright\scriptsize\tt#1}}
\def\draft{\oddsidemargin -.5truein
        \def\@oddfoot{\sl preliminary draft \hfil
        \rm\thepage\hfil\sl\today\quad\militarytime}
        \let\@evenfoot\@oddfoot \overfullrule 3pt
        \let\label=\draftlabel
        \let\marginnote=\draftmarginnote

\def\@eqnnum{(\theequation)\rlap{\kern\marginparsep\tt\@eqnlabel}%
\global\let\@eqnlabel\@vacuum}  }
\def\preprint{\twocolumn\sloppy\flushbottom\parindent 1em
        \leftmargini 2em\leftmarginv .5em\leftmarginvi .5em
        \oddsidemargin -.5in    \evensidemargin -.5in
        \columnsep 15mm \footheight 0pt
        \textwidth 250mmin      \topmargin  -.4in
        \headheight 12pt \topskip .4in
        \textheight 175mm
        \footskip 0pt

\def\@oddhead{\thepage\hfil\addtocounter{page}{1}\thepage}
        \let\@evenhead\@oddhead \def\@oddfoot{} \def\@evenfoot{}
}
\def\titlepage{\@restonecolfalse\if@twocolumn\@restonecoltrue\onecolumn
     \else \newpage \fi \thispagestyle{empty}\c@page\z@
        \def\thefootnote{\fnsymbol{footnote}} }
\def\endtitlepage{\if@restonecol\twocolumn \else  \fi
        \def\thefootnote{\arabic{footnote}}
        \setcounter{footnote}{0}}  %\c@footnote\z@ }
\catcode`@=12 \relax
\def\ps@headings{\def\@oddfoot{}\def\@evenfoot{}
\def\@oddhead{\hbox{}\hfill
        \makebox[.5\textwidth]{\raggedright\ignorespaces --\thepage{}--
        \hfill }}
\def\@evenhead{\@oddhead}
\def\subsectionmark##1{\markboth{##1}{}}
}

\ps@headings

\relax

\usepackage{graphicx}

\begin{document}
%%%%%%%%%%%%%%%%%%%%%%%%%%%%%%%%%%%%%%%%%%%%%%
%\begin{figure}[h]
%\centering
%\includegraphics[scale=0.5]{a:figure1p.ps}
%\end{figure}
%%%%%%%%%%%%%%%%%%%%%%%%%%%%%%%%%%%%%%%%%%%%%%

\begin{titlepage}
\begin{flushright}

NTUA--98--2000

hep-th/0010014 \\

\end{flushright}

\begin{centering}
\vspace{.41in}
{\large {\bf Cosmological Evolution of a Brane Universe in a Type
0 String Background.}}\\

\vspace{.2in}

{\bf E.~Papantonopoulos}$^{a}$ and {\bf I. Pappa} $^{b}$ \\
\vspace{.2in}

 National Technical University of Athens, Physics
Department, Zografou Campus,\\ GR 157 80, Athens, Greece. \\

\vspace{0.5in}

{\bf Abstract} \\

\end{centering}

\vspace{.1in}
We study the cosmological evolution of a D3-brane Universe in a
type 0 string background. We follow the brane-universe along the
radial coordinate of the background and we calculate the energy
density which is induced on the brane because of its motion in the
bulk. We find that for some typical values of the parameters and
for a particular range of values of the scale factor of the
brane-universe, the effective energy density is dominated by a
term proportional  to $\frac{1}{(loga)^{4}}$ indicating a slow
varying inflationary phase.  For larger values of the scale factor
the effective energy density takes a constant value and the
brane-universe enters its usual inflationary period.

\vspace{0.5in}
\begin{flushleft}

$^{a}$ e-mail address:lpapa@central.ntua.gr \\$ ^{b}$ e-mail
address:gpappa@central.ntua.gr

\end{flushleft}

\end{titlepage}

\section{Introduction}

 There has been much recent interest in the idea that our universe
 may be a brane embedded in some higher dimensional space
 \cite{Reg}. It has been shown that the hierarchy problem can be
 solved if the higher dimensional Planck scale is low and the
 extra dimensions large \cite{Dim}. Randall and Sundrum \cite{Rand}
 proposed a solution of the hierarchy problem
 without the need for large extra dimensions but instead through
 curved five-dimensional spacetime $AdS_{5}$ that generates an
 exponential suppression of scales.

 This idea of a brane-universe can  naturally be applied to string
 theory. In this context, the Standard Model gauge bosons as well as
 charged matter arise as fluctuations of the D-branes. The universe
 is living on a collection of coincident branes, while gravity and
 other universal interactions is living in the bulk space
 \cite{Pol}. For example, the strongly coupled $E_{8} \times E_{8}$
 heterotic string theory is believed to be an eleven-dimensional
 theory, the field theory limit of which was described by Horava
 and Witten \cite{Wit}. The spacetime manifold includes a compact
 dimension with an orbifold structure. Matter is confined on the
 two ten-dimensional hypersurfaces (nine-branes) which can be seen
 as forming the boundaries of this spacetime.

 This new concept of brane-universe naturally leads to a new
 approach to cosmology. Any cosmological evolution like inflation
 has to take place on the brane while gravity acts globally on
 the whole space. In the literature there are a lot of cosmological
 models which study the cosmological evolution of our universe. In
 most of these models the spacetime is five-dimensional, where
 the fifth dimension is the radial dimension of an $AdS_{5}$
 space. The effective Einstein equations on the brane are then solved
 taking under consideration the matter on the brane  \cite{Cosm}-\cite{Keh}.

 Another approach to cosmological evolution of our brane-universe
 is to consider the motion of the brane in  higher
 dimensional spacetimes. In \cite{Cha} the motion of a domain wall
 (brane) in such a space  was studied. The Israel matching conditions
 were used to relate the bulk to the domain wall (brane) metric, and
 some interesting cosmological solutions were found. In
 \cite{Keh} a universe three-brane is considered in motion
 in ten-dimensional space in the presence of a gravitational field
 of other branes. It was shown that this motion in ambient space
 induces cosmological expansion (or contraction) on our universe,
 simulating various kinds of matter.

 In this direction we have studied \cite{Papa}, the motion of a three-brane
 in the background of type 0 string theory . It was shown
 that the motion of the brane on this specific background,
 with constant values of dilaton and tachyon fields,
 induces a cosmological evolution which for some range of the
 parameters has an inflationary phase. In \cite{Kim},
 using similar technics  the
 cosmological evolution of the three-brane in the background of
 type IIB string theory was considered.

 Type 0 string theories \cite{Tset}
 are interesting because of their connection
 \cite{Typ0} to four-dimensional
 $SU(N)$ gauge theory. The type 0 string does not have spacetime
 supersymmetry and because of that contains in its spectrum a
 non-vanishing tachyon field. In \cite{Tset} it was argued that one
 could construct the dual of an SU(N) gauge theory with 6 real
 adjoint scalars by stacking N electric three-dimensional branes
 of the type 0 model on top of each other. The tachyon field
 couples to the five form field strength, which drives the tachyon
 to a nonzero expectation value.

 Asymptotic solutions of the dual gravity background were
 constructed in \cite{Tset,Minah}. At large radial coordinate
 the tachyon is constant and one
 finds a metric of the form $AdS_{5} \times S^{5}$ with vanishing
 coupling which was interpreted as a UV fixed point. The solution
 exhibits a logarithmic running in qualitative agreement with the
 asymptotic freedom property of the field theory. At small radial
 coordinate the tachyon vanishes and one finds again a solution of the
 form $AdS_{5} \times S^{5}$ with infinite coupling, which was
 interpreted as a strong coupling IR fixed point.
 A gravity solution which describes the flow from the UV
 fixed point to the IR fixed point is given in \cite{Magg}.

 We calculate the effective energy density which is induced
 on the brane because of its motion in the particular background
 of a type 0 string. Using the approximate solutions of
 {\cite{Tset,Minah,Magg}, we find that for large values of the
 radial coordinate $r$, in the UV region, the effective energy density
 takes a constant value, which means that the universe has an
 inflationary period. For smaller values of $r$, or of the scale factor
 $\alpha$, the energy  density is dominated by a term proportional to
 $ \frac{1}{(loga)^{4}}$, where $\alpha$ is the scale factor of the
 brane-universe. This value of the energy density indicates that
 the universe is in a slow inflationary phase,
 in a "logarithmic inflationary" phase as we can call it, in contrast to
 "constant inflationary" phase which characterizes the usual
 exponential behaviour.
 For even smaller values of $r$,
 the  approximation breaks down and we cannot trust anymore
 the solutions. If we go to the IR region the energy density is
 dominated by the term  $ \frac{1}{\alpha^{4}} $ and again
 we find the "logarithmic inflation" for larger values of $r$. The
 approximation breaks down again for some larger values of $r$. It is well
 known that it is very difficult to connect the IR to the UV
 solutions. Therefore our failure to present a full cosmological
 evolution, relies exactly on this fact \cite{papanto}.

 We note here that what we find is somewhat peculiar, in the sense that one
 does not expect the effective energy density to be dominated, for
 a range of values of the scale factor, by terms proportional to
 $ \frac{1}{(loga)^{4}}$.  We understand this behaviour, as due entirely
 to mirage matter which is induced on the brane, from this particular background.

 Our work is organized as follows. In section two, we
 develop the formalism for a brane moving in a string background
 with a dilaton and a RR field. In section three, we discuss
 type 0 string and except the exact solution in an $AdS_{5} \times S^{5}$
 background we discuss the asymptotic UV and IR solutions of type 0 strings.
 In section four, we discuss the cosmological evolution of a brane
 in the background of type 0 string. Finally in the last
 section  we discuss our results.

 \section{Brane moving in ten-dimensional background}

 We will consider a probe brane moving in a generic
 static, spherically symmetric background \cite{Keh}.
 The brane will move in a geodesic.
 We assume the brane to be light compared to the background so
 that we will neglect the back-reaction.
 As the brane moves
 the induced world-volume metric becomes a function of
 time, so there is a cosmological evolution from the brane point
 of view.
 The metric of a D3-brane is parametrized as
 \begin{equation}\label{in.met}
ds^{2}_{10}=g_{00}(r)dt^{2}+g(r)(d\vec{x})^{2}+
  g_{rr}(r)dr^{2}+g_{S}(r)d\Omega_{5}
\end{equation}
 and there is also a dilaton field $\Phi$ as well as a $RR$
 background~$C(r)=C_{0...3}(r)$ with a self-dual field strength.
 The dynamics on the brane will be governed by the
 Dirac-Born-Infeld action  given by

\begin{eqnarray}\label{B.I. action}
  S&=&T_{3}~\int~d^{4} \xi
  e^{-\Phi}\sqrt{-det(\hat{G}_{\alpha\beta}+(2\pi\alpha')F_{\alpha\beta}-
  B_{\alpha\beta})}
   \nonumber \\&&
  +T_{3}~\int~d^{4} \xi\hat{C}_{4}+anomaly~terms
\end{eqnarray}
 The induced metric on the brane is
\begin{equation}\label{ind.metric}
  \hat{G}_{\alpha\beta}=G_{\mu\nu}\frac{\partial x^{\mu}\partial x^{\nu}}
  {\partial \xi^{\alpha}\partial\xi^{\beta}}
\end{equation}
 with similar expressions for $F_{\alpha\beta}$ and
 $B_{\alpha\beta}$. In the static
 gauge, $x^{\alpha}=\xi^{\alpha},\alpha=0,1,2,3 $
 using (\ref{ind.metric}) we can calculate the bosonic part of the
 brane Lagrangian which reads
\begin{equation}\label{brane Lagr}
L=\sqrt{A(r)-B(r)\dot{r}^{2}-D(r)h_{ij}\dot{\varphi}^{i}\dot{\varphi}^{j}}
-C(r)
\end{equation}
where $h_{ij}d \varphi ^{i} d \varphi^{j}$ is the line
 element of the unit five-sphere, and
\begin{equation}\label{met.fun}
  A(r)=g^{3}(r)|g_{00}(r)|e^{-2\Phi},
  B(r)=g^{3}(r)g_{rr}(r)e^{-2\Phi},
  D(r)=g^{3}(r)g_{S}(r)e^{-2\Phi}
\end{equation}
and $C(r)$ is the $RR$ background. Demanding conservation of
energy $E$ and of total angular
 momentum $ \ell ^{2} $ on the brane,
we find

\begin{equation}\label{functions}
\dot{r}^{2}=\frac{A}{B}(1-\frac{A}{(C+E)^{2}}\frac{D+\ell^{2}}{D}),
h_{ij}\dot{\varphi}^{i}\dot{\varphi}^{j}=\frac{A^{2}\ell^{2}}{D^{2}(C+E)^{2}}
\end{equation}
We can see that the above relation gives the following constraint
\begin{equation}\label{constr}
  (1-\frac{A}{(C+E)^{2}}\frac{D+\ell^{2}}{D})\geq 0
\end{equation}
The induced four-dimensional metric
 on the brane is

\begin{equation}\label{fmet}
d\hat{s}^{2}=(g_{00}+g_{rr}\dot{r}^{2}+g_{S}h_{ij}\dot{\varphi}^{i}\dot{\varphi}^{j})dt^{2}
+g(d\vec{x})^{2}
\end{equation}
In the above relation we substitute $ \dot{r}^{2}$ and $
h_{ij}\dot{\varphi}^{i}\dot{\varphi}^{j} $ from (\ref{functions}),
and we get

\begin{equation}\label{fin.ind.metric}
d\hat{s}^{2}=-d\eta^{2}+g(r(\eta))(d\vec{x})^{2}
\end{equation}
 with $\eta$ the cosmic time which is defined  by
\begin{equation}\label{cosmic}
 d\eta=\frac{|g_{00}|g^{\frac{3}{2}}e^{-\Phi}}{|C+E|}dt
\end{equation}
The induced metric (\ref{fin.ind.metric}) on the brane, is the
standard form of a flat expanding universe. For this metric we can
write the effective Einstein equations on the brane,
\begin{equation}\label{Einst1}
  R_{\mu \nu}-\frac{1}{2} g_{\mu \nu} R =8\pi G (T_{\mu \nu})_{eff}
\end{equation}
where $(T_{\mu\nu})_{eff}$ is the effective energy momentum tensor
which is induced on the brane. We have assumed that our brane is
light and there is no back-reaction with the bulk. We expect  the
$(T_{\mu \nu})_{eff}$ to be a function of the quantities of the
bulk.

Before we proceed, it is interesting to discuss the general case
where there is matter on the brane with an energy momentum tensor
$\tau_{\mu\nu}$ and a cosmological constant $\Lambda$. The
Einstein equations on the brane are \cite{mala1}, \cite{mala2}
\begin{equation}\label{mal1}
R_{\mu\nu}-\frac{1}{2} g_{\mu\nu} R =-\Lambda q_{\mu\nu} +8\pi G
\tau_{\mu\nu} +\kappa^{4}\pi_{\mu\nu} - E_{\mu\nu}
\end{equation}
where $q_{\mu\nu}$  is the induced metric on the brane,
$\pi_{\mu\nu}$ is a function of the matter content of the brane,
having the form
\begin{equation}\label{mal2}
\pi_{\mu\nu} = -\frac{1}{4}\tau_{\mu\nu} \tau_{\nu}^{\alpha}
+\frac{1}{12} \tau \tau_{\mu\nu} +\frac{1}{8}
q_{\mu\nu}\tau_{\alpha\beta}\tau^{\alpha\beta} - \frac{1}{24}
q_{\mu\nu}\tau^{2}
\end{equation}
and $E_{\mu\nu}$ is given by
\begin{equation}\label{tensor}
E_{\mu\nu}=C^{\alpha}_{\beta\rho\sigma}n_{\alpha}n^{\rho}q_{\mu}
^{\beta}q_{\nu}^{\sigma}
\end{equation}
where $C^{\alpha}_{\beta\rho\sigma}$ is the Weyl tensor
\cite{mala1}. As we can see from the above relation the term
$E_{\mu\nu}$ is a geometrical object depending on the bulk
geometry. If we now compare (\ref{Einst1}) with (\ref{mal1}), we
can see that because in our case we do not have matter on the
brane, consistency requires that $(T_{\mu\nu})_{eff}$ of equation
(\ref{Einst1}) should be proportional to $E_{\mu\nu}$ of
(\ref{mal1}). One can check that using relation (\ref{tensor}) one
can get similar results as ours, an approach which is followed in
\cite{kraus}.

If we now assume the usual form of a perfect fluid for the
effective energy momentum tensor, we get from  (\ref{Einst1})
\begin{equation} \label{r-tt}
 8 \pi G \rho +\Lambda= \frac{3}{4} g^{-2}\dot{g}^{2}
 \end{equation}
We can now define an $\rho_{eff}$ from the relation
 \begin{equation}\label{ident}
  8 \pi G \rho +\Lambda= 8 \pi G_{N} \rho_{eff}
 \end{equation}
Using equation (\ref{cosmic}) we get
\begin{equation} \label{gprim}
 \dot{g}=g^{\prime}\Big{[} \frac {|g_{00}|}{g_{rr}}
 -\frac {g_{00}^{2}}{g_{s} g_{rr}} \Big{(}
 \frac{ g^{3}g_{s}e^{-2\Phi}+\ell^{2} }
 {(C+E)^{2}} \Big{)}
 \Big{]} ^{\frac{1}{2}}
  \frac{|C+E|}
  {|g_{00}|g^{\frac{3}{2}} e^{-\Phi}}
\end{equation}
where prime denotes differentiation with respect to $r$.
 To derive an analogue of the four dimensional Friedman equations
 for the expanding four dimensional universe on the probe
 D3-brane, we define the scale factor as $g=\alpha^{2}$ and then equation
(\ref{ident}) with the use of  (\ref{r-tt}) and (\ref{gprim})
becomes
\begin{equation}\label{dens}
 \frac{8\pi}{3}G_{N}\rho_{eff}= (\frac
{\dot{\alpha}}{\alpha})^{2}=
\frac{(C+E)^{2}g_{S}e^{2\Phi}-|g_{00}|(g_{S}g^{3}+\ell^{2}e^{2\Phi})}
{4|g_{00}|g_{rr}g_{S}g^{3}}(\frac{g'}{g})^{2}
\end{equation}

Therefore the motion of a D3-brane on a general spherically
symmetric background had induced on the brane a matter density. As
it is obvious from the above relation, the specific form of the
background will determine the cosmological evolution on the brane.

We had defined $\rho_{eff}$ through the relation (\ref{ident}). In
this relation, the 4-dimensional Newton's constant $G_{N}$ and the
four dimensional cosmological constant appears. To make our
approach clear, we will discuss their physical meaning in our
scheme.

We have assumed that in our brane-universe there is no gravity by
itself therefore the Newton's law is defined on the whole
10-dimensional space. As the brane moves, we can write on the
brane the effective Einstein equations (\ref{Einst1}). The
Newton's constant which appears in this equation has the meaning
of an effective parameter determined by the background. To see its
value let us go to a particular background. Let us consider the
near-horizon geometry of a macroscopic D3-brane which is
$AdS_{5}\times S^{5}$, with metric \cite{Keh,kraus},

\begin{equation}\label{gianna1}
ds^{2} =
\frac{r^{2}}{L^{2}}\Big{[}-\Big{(}1-(\frac{r_{0}}{r})^{4}\Big{)}dt^{2}
+(d\overrightarrow{x})^{2}\Big{]}
+\frac{L^{2}}{r^{2}}\frac{dr^{2}}{\Big{(}1-(\frac{r_{0}}{r})^{4}\Big{)}}
+L^{2}d\Omega_{5}^{2}
\end{equation}
where $L^{4}=4\pi g N \alpha'^{2}$ with g the string coupling in
10 dimensions and N the number of D3-branes. By using (\ref{dens})
the effective energy density on the brane in this background is
\begin{equation}\label{gianna2}
\frac{8\pi}{3}G_{N}\rho_{eff} =
\frac{1}{L^{2}}\Big{[}(1+\frac{E}{\alpha^{4}})^{2}-\Big{(}1-(\frac{r_{0}}{L})^{4}
\frac{1}{\alpha^{4}}\Big{)}
\Big{(}1+\frac{\ell^{2}}{L^{2}}\frac{1}{\alpha^{6}}\Big{)}\Big{]}
\end{equation}
As we can see, the only scale which enters in this relation is
$\alpha'$. Then as it is obvious from the above relation, we can
on purely dimensional grounds write $G_{N}=L^{2}$ or $G_{N} = 2
\sqrt{\pi g N}\alpha'$. We can express the $G_{N}$ in terms of the
5-dimensional Newton's constant using the relation
$T_{3}=\frac{3}{4\pi G L}$ \cite{kraus}
\begin{equation}\label{gianna3}
G_{N} = \frac{4\pi}{3}G T_{3} (2\sqrt{\pi g
N}\alpha')^{\frac{3}{2}}
\end{equation}
where $T_{3}$ is the brane tension.

 The induced cosmological constant on the brane $\Lambda$ can
be expressed in terms of the background fields using the equation
\begin{equation}\label{gianna4}
-g^{-2}\dot{g}^{2}+g^{-1}\ddot{g}
 +\frac{3}{4}\gamma g^{-2}\dot{g}^{2}
 -\gamma\Lambda = 0
\end{equation}
In deriving the above equation we have used the conservation of
energy momentum tensor $(T_{\mu\nu})_{eff}$ and an equation of
state in the form $p=(\gamma-1)\rho$. Nevertheless we believe that
in this approach we cannot really distinguish between $\rho$ and
$\Lambda$ and a definition of the form (\ref{ident}) is more
meaningful. As it is discussed in \cite{mala1} in the case where
we cannot distinguish between vacuum energy and matter energy we
cannot truly specify $G_{N}$. Then as we discussed above, $G_{N}$
is determined in a "phenomenological" way depending on the
background.

\section{Type 0 string background}

Type 0 string theory is interesting because of its connection to
gauge theories. This enables us to study SU(N) gauge theory by
merely gravitational quantities. Another advantage of this theory
is the presence of a tachyon field. Tachyonic fields in ordinary
field theory create instabilities. In cosmology on the contrary,
the time evolution of a tachyon field plays an important
$r\hat{o}le$. In two dimensions because the tachyon field is a
matter field has important consequences in cosmology \cite{Dia},
and it can give a solution to the "gracefull exit" problem
\cite{Pap}. In four dimensions its effect to cosmology has been
examined by various authors \cite{Perry}.

As we have shown in \cite{Papa} in type 0 string the tachyon field
can induce inflation on the brane. We had used an exact solution
of type 0 string with constant tachyon and dilaton fields. If
these fields are coordinate dependent then, there is not an exact
solution of the theory, but there are approximate solutions which
we will discuss in the following.
 The action of the type 0 string is given
by \cite{Tset}
\begin{eqnarray}\label{action}
S_{10}&=&~\int~d^{10}x\sqrt{-g}\Big{[} e^{-2\Phi} \Big{(}
 R+4(\partial_{\mu}\Phi)^{2} -\frac{1}{4}(\partial_{\mu}T)^{2}
-\frac{1}{4}m^{2}T^{2}-\frac{1}{12}H_{\mu\nu\rho}H^{\mu\nu\rho}\Big{)}
\nonumber \\&& - \frac{1}{4}(1+T+\frac{T^{2}}{2})|F_{5}|^{2}
\Big{]}
\end{eqnarray}

The equations of motion which result from this action, with the
antisymmetric field put to zero, are
\begin{equation}\label{dilaton}
  2\nabla^{2}\Phi-4(\nabla_{n}\Phi)^{2}-\frac{1}{4}m^{2}T^{2}=0
\end{equation}

\begin{eqnarray}\label{metric}
  &R_{mn}&+2\nabla_{m}\nabla_{n}\Phi-\frac{1}{4}\nabla_{m}T\nabla_{n}T
  -\frac{1}{4\cdot4!}e^{2\Phi}f(T) \Big{(}F_{mklpq}F_{n}~^{klpq}\nonumber
  \\&&  - \frac{1}{10}G_{mn}F_{sklpq}F^{sklpq} \Big{)}=0
\end{eqnarray}

\begin{equation}\label{Tachyon}
  (-\nabla^{2}+2\nabla^{n}\Phi\nabla_{n}+m^{2})T
  +\frac{1}{2\cdot5!}e^{2\Phi}f'(T)F_{sklpq}F^{sklpq}=0
\end{equation}

\begin{equation}\label{F}
  \nabla_{m} \Big{(}f(T)F^{mnkpq} \Big{)}=0
\end{equation}
The tachyon is coupled to the $RR$ field through the function
\begin{equation}\label{ftac}
 f(T)=1+T+\frac{1}{2} T^{2}
\end{equation}
In the background where the tachyon field acquires vacuum
expectation value $T_{vac}=T_{0}=-1$, the tachyon function
(\ref{ftac}) takes the value $f(T_{0})=\frac{1}{2}$ which
guarantee the stability of the theory \cite{Tset}.

The equations (\ref{dilaton})-(\ref{F}) can be solved using the
following ansatz for the metric

\begin{eqnarray}\label{tenmetric}
ds^{2}_{10}=g_{00}(r)dt^{2}+g(r)(d\vec{x})^{2}+
  g_{rr}(r)dr^{2}+g_{S}(r)d\Omega_{5}
\end{eqnarray}
Technically it is easier to solve the above equations if we go to
new variables. One can then introduce new parameter $ \rho $
through the relation

\begin{equation}\label{rho}
\rho = \frac{e^{2\Phi_{0}}}{4r^{4}}
\end{equation}

and the fields $ \xi $ and $\eta$ from the relations

\begin{equation}\label{xxi}
g = e^{\frac{\Phi-\xi}{2}}
\end{equation}

\begin{equation}\label{eta}
g_{s} = e^{\frac{\Phi+\xi}{2}-\eta}
\end{equation}
Then (\ref{tenmetric}) takes the form,

\begin{equation}
ds^{2} = - e^{\frac{\Phi-\xi}{2}}dt^{2}+
e^{\frac{\Phi-\xi}{2}}d\vec{x}^{2}+e^{\frac{\Phi+\xi}{2}-5\eta}d\rho^{2}
 +e^{\frac{\Phi+\xi}{2}-\eta}d\Omega_{5}^{2}
\end{equation}

 With this form of the metric the action (\ref{action}) can be described by the
following Toda-like mechanical system (dot denotes
$\rho$-derivative)

\begin{equation}\label{Tota}
S = \int d\rho [\frac{1}{2}\dot{\Phi}^{2}
+\frac{1}{2}\dot{\xi}^{2} +\frac{\dot{T}^{2}}{4} -5\dot{\eta}^{2}
-V(\Phi,\xi,T,\eta)]
\end{equation}
with the potential $ V(\Phi,\xi,T,\eta)$ given by

\begin{equation}\label{Poten}
V(\Phi,\xi,T,\eta) = g(T)e^{\frac{1}{2}\Phi + \frac{1}{2}\xi
-5\eta} +20 e^{-4\eta} -Q^{2}f^{-1}(T)e^{-2\xi}
\end{equation}

 If the tachyon field takes its vacuum value and the dilaton
field a constant value $\Phi=\Phi_{0}$ one can find the
electrically charged three-brane

\begin{equation}\label{Sol}
  g_{00}=-H^{-\frac{1}{2}},
  g(r)=H^{-\frac{1}{2}},  g_{S}(r)=H^{\frac{1}{2}}r^{2},
g_{rr}(r)=H^{\frac{1}{2}},    H=1+\frac{e^{\Phi_{0}}Q}{2r^{4}}
\end{equation}
if the following ansatz for the $RR$ field

\begin{equation}\label{Form}
  C_{0123}=A(r),   F_{0123r}=A'(r)
\end{equation}
is used.

If $T$ and $\Phi$ are functions of the coordinate $r$, then
approximate solutions exist \cite{Tset,Minah}. These solutions are
valid for large (UV) and small (IR) values of the radial
coordinate. The approximations of \cite{Tset,Minah} agree in the
UV  region but in the IR the approximation of  \cite{Tset} leads
to an IR fixed point, while the approximation of \cite{Minah} to a
confining point.

From the action (\ref{Tota}) we can derive the following equations
of motion \cite{Minah}

\begin{equation}\label{eqxi}
\ddot{\xi} +\frac{1}{2}g(T)e^{\frac{1}{2}\Phi + \frac{1}{2}\xi
-5\eta} +2\frac{Q^{2}}{f(T)}e^{-2\xi}=0
\end{equation}

\begin{equation}\label{eqeta}
\ddot{\eta} + \frac{1}{2}g(T)e^{\frac{1}{2}\Phi + \frac{1}{2}\xi
-5\eta} +8e^{-4\eta} =0
\end{equation}

\begin{equation}\label{eqphi}
\ddot{\Phi}+\frac{1}{2}g(T)e^{\frac{1}{2}\Phi + \frac{1}{2}\xi
-5\eta} =0
\end{equation}

\begin{equation}\label{eqT}
\ddot{T} +2g'(T)e^{\frac{1}{2}\Phi + \frac{1}{2}\xi -5\eta}
+2\frac{Q^{2}f'(T)}{f^{2}(T)}e^{-2\xi} = 0
\end{equation}

where $g(T)$ is the bare tachyon potential,
\begin{equation}\label{gfun}
g(T)=\frac{1}{2} T^{2}- \lambda T^{4}
\end{equation}
and $\lambda$ is a parameter. Defining a new variable
$\rho=u^{-4}$, in the UV for $u\longrightarrow\infty$, or
$\rho\longrightarrow 0$,
 we can solve the equations of motion
(\ref{eqxi})-(\ref{eqT}) to the next to leading order and  find
\cite{Minah,Tset}

\begin{equation}\label{apprT}
T = T_{0}-4\frac{g'(T_{0})}{g(T_{0})}\frac{1}{log \rho} +
O(\frac{log(-log\rho)}{log^{2}\rho})
\end{equation}

\begin{equation}\label{apprphi}
\Phi = -2log(C_{0}log\rho) -
\Big{(}7+8(\frac{g'(T_{0})}{g(T_{0})})^{2}\Big{)}
\Big{(}\frac{log(-log\rho)}{log\rho}\Big{)} +\frac{B}{log\rho} +
O(\frac{log^{2}(-log\rho)}{log^{2}\rho})
\end{equation}

\begin{equation}\label{apprxi}
\xi = log\Big{(}\sqrt{2f^{-1}(T_{0})}Q\rho\Big{)} -
\frac{1}{log\rho} + O(\frac{log(-log\rho)}{log^{2}\rho})
\end{equation}

\begin{equation}\label{appreta}
\eta = \frac{1}{2}log(4\rho)- \frac{1}{log\rho} +
O(\frac{log(-log\rho)}{log^{2}\rho})
\end{equation}

 where $C_{0}=- \frac{4C_{2}^{5}} {g(T_{0}) \surd C_{1}} $ and

\begin{equation}\label{Cfun}
C_{1} =
\frac{2Q}{\sqrt{2f(T_{0})}}(1+\frac{1}{4log\frac{u}{u_{0}}}) ,
C_{2} = 2(1+\frac{1}{4log\frac{u}{u_{0}}})
\end{equation}

The above solutions show that at the UV point, the tachyon takes a
constant value and if we calculate the next to leading order
effective coupling $e^{\frac{1}{2}\Phi}$ ,

\begin{equation}\label{coupl}
e^{\frac{1}{2}\Phi} \sim
\frac{1}{logu-(\frac{7}{8}+\frac{g'(T_{0})^{2}}{g(T_{0})^{2}})loglogu}
\end{equation}

we see that goes to zero for large $u$.

For $u\longrightarrow 0$ or for large $\rho$ there are two
approximate solutions in the literature, leading to infrared fixed
point \cite{Tset} or to a confining fixed point \cite{Minah,Magg}.
For the first approximation we have

\begin{equation}\label{tset1}
T = -\frac{16}{log\rho}- \frac{8}{log^{2}\rho}(9loglog\rho-3) +
O(\frac{log^{2}log\rho}{log^{3}\rho})
\end{equation}

\begin{equation}\label{tset2}
\Phi = -\frac{1}{2}log(2Q^{2}) +2loglog\rho -
\frac{1}{log\rho}9loglog\rho + O(\frac{loglog\rho}{log^{2}\rho})
\end{equation}

\begin{equation}\label{tset3}
\xi = \frac{1}{2}log(2Q^{2}) + log\rho + \frac{9}{log\rho}
+\frac{9}{2log^{2}\rho}(9loglog\rho-\frac{20}{9}) +
O(\frac{log^{2}log\rho}{log^{3}\rho})
\end{equation}

\begin{equation}\label{tset4}
\eta = log2 + \frac{1}{2}log\rho + \frac{1}{log\rho}
+\frac{1}{2log^{2}\rho}(9loglog\rho-2) +
O(\frac{log^{2}log\rho}{log^{3}\rho})
\end{equation}

While for the second solution  \cite{Minah,Magg} we have

\begin{equation}\label{min1}
\Phi =
\Phi_{0}+\rho-\frac{1}{16(\sqrt{5}+3)^{2}}e^{\frac{\Phi_{0}}{2}}e^{-(\sqrt{5}+3)\rho}
\end{equation}

\begin{equation}\label{min2}
\eta =
\frac{1}{\sqrt{5}}\rho-\frac{5}{2}e^{-\frac{4}{\sqrt{5}}\rho}
\end{equation}

\begin{equation}\label{min3}
\xi = \rho-\frac{1}{2}e^{-2\rho}
\end{equation}

\begin{equation}\label{min4}
T = -\frac{1}{2}e^{-2\rho}
\end{equation}

In both approximations the tachyon field in the IR point goes to
zero while the effective coupling gets infinite. It is important
to observe that the approximate solutions in the UV
(\ref{apprT})-(\ref{appreta})  and in the IR
(\ref{tset2})-(\ref{tset4}) of \cite{Tset} are related by $ y
\longrightarrow -y$, suggesting that they can be smoothly
connected into a full interpolating solution. An attempt to
connect these solutions was presented in \cite{Magg}. As we
mention above, the tachyon field starts at T=-1 at $\rho=0$ in the
UV , and grows according to (\ref{apprT}), then enters an
oscillating regime and finally relaxes to zero according to
(\ref{tset1}), when $\rho=\infty$ in IR. This guarantees that
$T^{2}e^{\frac{1}{2}\Phi}$ becomes small which leads the metric in
the $AdS_{5} \times S^{5}$ form.

There is a question if we can trust the asymptotic solutions in
the infrared. The problem is that when the coupling becomes
strong, string corrections become important. The situation is
different in the UV where we can trust our solutions because the
coupling is small. The role of the $\alpha^{'}$ corrections in the
IR has been discussed in \cite{Minah,Magg}. It was claimed that
the $\alpha^{'}$ corrections are small.

\section{Cosmological evolution of the Brane-Universe }

We consider a D3-brane moving along a geodesic in the background
of a type 0 string. Having all the solutions in the ultra violet
and the infrared, we can follow the cosmological evolution of our
universe as it moves along the radial coordinate $r$. In the
presence of a non trivial tachyon field the coupling $e^{-\Phi}$
which appears in the Dirac-Born-Infeld action in (\ref{B.I.
action}), is modified by a tachyonic function
$\kappa(T)=1+\frac{1}{4}T+O(T^{2})$. Then we can define an
effective coupling \cite{Typ0}
\begin{equation}\label{effphi}
e^{-\Phi}_{eff}=\kappa(T) e^{-\Phi}
\end{equation}

 The bulk
fields are also coordinate dependent and the induced metric on the
brane  will depend on a non trivial way on the dilaton field.
Therefore the metric in the string frame will be connected to the
metric in the Einstein frame through
$g_{St}=e^{\frac{\Phi}{2}}_{eff} g_{E}$.  All the quantities used
so far were defined in the string frame. We will follow our
cosmological evolution in the Einstein frame. Then the relation
(\ref{dens}) becomes

\begin{equation}\label{eindens}
 \frac{8\pi}{3}\rho_{eff}= (\frac {\dot{\alpha}}{\alpha})^{2}=
\frac{(C+E)^{2}g_{S}-|g_{00}|(g_{S}g^{3}+\ell^{2})}
{4|g_{00}|g_{rr}g_{S}g^{3}}(\frac{g'}{g})^{2}
\end{equation}

Having the approximate solution in the UV given by
(\ref{apprT})-(\ref{appreta}) we can calculate the metric
components of the metric (\ref{tenmetric}) and find

\begin{equation}\label{gyy}
g_{yy} = \frac{1}{16}\sqrt{\frac{Q}{2}}(1-\frac{9}{2y})
\end{equation}

\begin{equation}\label{g}
g = \frac{1}{\sqrt{2Q}}e^{\frac{y}{2}}(1-\frac{1}{2y})
\end{equation}

\begin{equation}\label{gs}
g_{s} = \sqrt{\frac{Q}{2}}(1-\frac{1}{2y})
\end{equation}

The variable y is defined by

\begin{equation}
\rho = e^{-y}
\end{equation}

 Then we can identify $g$ of (\ref{g}) with the scale factor $\alpha^{2}$
 and solve for $y$. We get two solutions

\begin{equation}\label{y1}
y_{1} = -\frac{1}{4log\alpha + log2Q}
\end{equation}

\begin{equation}\label{y2}
y_{2} = log2Q + 4log\alpha + \frac{1}{log2Q + 4log\alpha}
\end{equation}

From the solution (\ref{y2}) which has the right behaviour for
large $\alpha$, we keep the $log2Q + 4log\alpha$ term. Then the
$RR$ field C becomes

\begin{equation}\label{cfield}
C = \frac{e^{y}}{Q} - \frac{2}{Q}Ei[y]
\end{equation}

Then, we can calculate the effective energy density from
(\ref{eindens}) setting $\ell^{2}=0$ and we get

\begin{eqnarray}\label{rhonimah}
\frac{8\pi}{3}\rho_{eff}& =&
 \Big{[} \Big{(}1-\frac{1}{Q\alpha^{4}}Ei[log2Q + 4log\alpha] +
\frac{E} {2\alpha^{4}} \Big{)}^{2} \nonumber \\ && -
\frac{1}{4}\Big{(}1-\frac{1}{2(log2Q + 4log\alpha)}\Big{)}^{4}
\Big{ ]}\Big{(}1-\frac{1}{2(log2Q +
4log\alpha)}\Big{)}^{-4}\Big{(}1-\frac{9}{2(log2Q +
4log\alpha)}\Big{)}^{-1} \nonumber \\&& \Big{(}1+\frac{1}{(log2Q +
4log\alpha)^{2}}\frac{1}{\Big{(}1-\frac{1}{2(log2Q +
4log\alpha)}\Big{)}}\Big{)}^{2}
\end{eqnarray}

For some typical value of the parameters Q=1 and E=1, and for
large values of $\alpha$, it is obvious that $\rho_{eff}$ has a
constant value. Therefore an observer on the brane will see an
expanding inflating universe. It is interesting to see what
happens for small values of $\alpha$. As $\alpha$ gets smaller, a
term proportional to $\frac{1}{(\log\alpha)^{4}}$ starts to
contribute to $\rho_{eff}$. Therefore the universe for small
values of scale factor has a slow expanding inflationary phase
which we call it "logarithmic inflationary" phase. For smaller
value of $\alpha$ we cannot trust the solution which is reflected
in the fact that $\rho_{eff}$ gets infinite. The behaviour of the
effective energy density as a function of the scale factor is
shown in Figure 1.

%%%%%%%%%%%%%%%%%%%%%%%%%%%%%%%%%%%%%%%%%%%%%%
%\begin{figure}[h]
%\centering
%\includegraphics[scale=0.7]{a:figure1a.ps}
%\centerline{\hbox{\psfig{figure1=figure1.eps,height=10cm}}}
%\caption {The induced energy density on the brane as a function of
%the brane scale factor.}
%\end{figure}
%%%%%%%%%%%%%%%%%%%%%%%%%%%%%%%%%%%%%%%%%%%%%%

%%%%%%%%%%%%%%%%%%%%%%%%%%%%%%%%%%%%%%%%%%%%%%
\begin{figure}[h]
\centering
\includegraphics[scale=0.9]{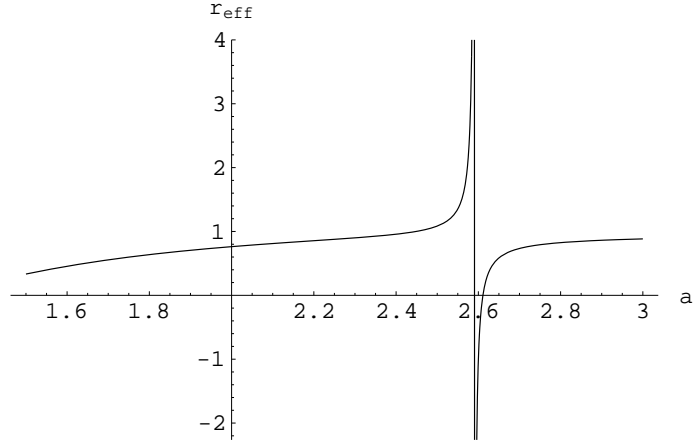}
%\centerline{\hbox{\psfig{figure1b=figure1b.eps,height=10cm}}}
\caption {The induced energy density on the brane as a function of
the brane scale factor.}
\end{figure}
%%%%%%%%%%%%%%%%%%%%%%%%%%%%%%%%%%%%%%%%%%%%%%

To have an idea how the slow inflationary phase proceeds, we can
assume for the moment that the effective energy density scales as
\begin{equation}
 \frac{8\pi}{3}\rho_{eff}=(\frac{\dot{\alpha}}{\alpha})^{2}
 =\frac{1}{(\log\alpha)^{p}}
 \end{equation}
The solution of the above equation is
\begin{equation}
\alpha=e^{ \textstyle  t^{ \textstyle (\frac{2}{p+2}) } }
\end{equation}

Therefore we remain in an exponentially growing universe, but
various values of p have the effect of making the universe to slow
down its expansion. We note here that in order to estimate the
behaviour and the duration of this "logarithmic inflationary"
phase, we have to resolve the problem of the singularity.

 Going now to IR using (\ref{tset1})-(\ref{tset4}) we get for
the metric components

\begin{equation}\label{tsegyy}
g_{yy} = \frac{\sqrt{Q}}{16}2^{-\frac{3}{4}}(1-\frac{1}{2y})
\end{equation}

\begin{equation}\label{tseg}
g =
\frac{2^{-\frac{1}{4}}}{\sqrt{Q}}e^{-\frac{y}{2}}(1-\frac{9}{2y})
\end{equation}

\begin{equation}\label{tsegs}
g_{s} = 2^{-\frac{3}{4}} \sqrt{Q}(1+\frac{7}{2y})
\end{equation}

where now y is defined by

\begin{equation}
\rho = e^{y}
\end{equation}

Then the identification $g=\alpha^{2}$ using (\ref{tseg}) gives
again two solutions

\begin{equation}\label{tsey1}
y_{1} = -\frac{9}{4log\alpha + log\sqrt{2}Q}
\end{equation}

\begin{equation}\label{tsey2}
y_{2} = -log\sqrt{2}Q - 4log\alpha + \frac{9}{log\sqrt{2}Q +
4log\alpha}
\end{equation}

 For small $\alpha$ we keep from the solution $y_{2}$ of (\ref{tsey2})
 the term $ -log\sqrt{2}Q - 4log\alpha$. Using this solution we can
 calculate the $RR$ field

\begin{equation}\label{ctset}
C =- \frac{e^{-y}}{Q} - \frac{2}{Q}Ei[-y]
\end{equation}

Then $\rho_{eff}$ becomes,

\begin{eqnarray}\label{tserho}
\frac{8\pi}{3}\rho_{eff}& =& \Big{[} \Big{
(}-1-2\frac{1}{\sqrt{2}Q\alpha^{4}}Ei[log\sqrt{2}Q + 4log\alpha] +
\frac{E}{\sqrt{2}\alpha^{4}} \Big{)}^{2} - \nonumber
\\ &&
 \frac{1}{2}\Big{(}1+\frac{9}{2(log\sqrt{2}Q +
4log\alpha)}\Big{)}^{4} \Big{]}\Big{(}1+ \frac{9}{2(log\sqrt{2}Q +
4log\alpha)}\Big{)}^{-4} \nonumber \\&&
\Big{(}1+\frac{1}{2(log\sqrt{2}Q + 4log\alpha)}\Big{)}^{-1}
\nonumber
\\&&
\Big{(}1-\frac{9}{2(4log\alpha+ log\sqrt{2}Q)^{2}}
 \frac{1}{\Big{(}1-\frac{9}{2(log\sqrt{2}Q + 4log\alpha)}\Big{)}}\Big{)}^{2}
\end{eqnarray}

As we can see, the above relation is the same as the energy
density in the UV (relation (\ref{rhonimah})) up to some numerical
factors, as expected. The difference is, that now it is valid for
small $\alpha$. For small $\alpha$ first the term $
\frac{1}{\alpha^{8}}$ dominates and then the term
$\frac{1}{\alpha^{4}}$. As $\alpha$ increases the term $
\frac{1}{(log\alpha)^{4}}$ takes over and drives the universe to a
slow inflationary expansion.

We will also discuss the cosmological behaviour of the  IR
solutions of \cite{Minah} and \cite{Magg}. Using
(\ref{min1})-(\ref{min4}) we
 have for the metric elements

\begin{equation}\label{mingyy}
g_{\rho\rho} =
e^{(\frac{1}{2}-\sqrt{5})\rho}e^{-\frac{1}{4}e^{-2\rho}+\frac{25}{2}
e^{-\frac{4\rho}{\sqrt{5}}}}\sqrt{Q}
\end{equation}

\begin{equation}\label{mingg}
g = e^{-\frac{\rho}{2}}e^{\frac{1}{4}e^{-2\rho}}\frac{1}{\sqrt{Q}}
\end{equation}

\begin{equation}\label{mings}
 g_{s} =
e^{(\frac{1}{2}-\frac{1}{\sqrt{5}})\rho}e^{-\frac{1}{4}
e^{-2\rho}+\frac{5}{2}e^{-\frac{4\rho}{\sqrt{5}}}}\sqrt{Q}
\end{equation}

 Then the equation $g=\alpha^{2}$ gives

\begin{equation}\label{miny}
\rho = -4log\alpha-logQ
\end{equation}

and the $RR$ field becomes

\begin{equation}\label{cmin}
C = (-2e^{-2\rho}-e^{-4\rho})\frac{1}{Q}
\end{equation}

Finally  $ \rho_{eff}$ becomes

\begin{eqnarray}\label{minirrho}
\frac{8\pi}{3}\rho_{eff}& =& (Q
\alpha^{4})^{-\sqrt{5}+\frac{1}{2}}\Big{[}(-2Q \alpha^{4}-Q^{3}
\alpha^{12}+E \alpha^{-4})^{2}-(1+Q^{2} \alpha^{8})\Big{]}
\nonumber
\\&&
\Big{(}1-\frac{1}{4}Q^{2}\alpha^{8}
+\frac{25}{2}(Q\alpha^{4})^{\frac{4}{\sqrt{5}}}\Big{)}^{-1}
\Big{(}1+Q^{2}\alpha^{8}\Big{)}
\end{eqnarray}

The above calculated effective energy density, in spite of its
different form, has a similar behaviour as (\ref{tserho}). As $
\alpha $ increases, various negative powers of $\alpha$ take over
until the singularity is reached where positive powers of $\alpha$
dominate.

\section{Discussion}

We had followed a probe brane along a geodesic in the background
of type 0 string. Assuming that the universe is described by a
three-dimensional brane, we calculate the effective energy density
which is induced on the brane because of this motion. We study
this mirage matter as the brane-universe moves along the radial
coordinate.

In our previous work \cite{Papa} we found, that the motion of the
brane-universe in this particular background induces an
inflationary phase on the brane. We made the analysis in the
limited case where the dilaton and tachyon fields were constants.
This assumption simplified the calculation because there is an
exact solution of the equations of motion.

In this work we extend our study to a background where all the
fields are functions of the radial coordinate. Then the problem
becomes more complicated because there is no more an exact
solution to the equations of motion. Nevertheless there are
approximate solutions for large values of the radial coordinate,
in the UV region and solutions for small values of the radial
coordinate in the IR. In the UV the coupling of the theory is
small, so we can trust the approximate solutions. In the IR, the
coupling becomes strong but it was shown in the literature
\cite{Minah,Magg}, that all string corrections are small.

Using these solutions, we calculate the energy densities that are
induced on the brane. What we find is that for large values of the
scale factor as it is measured on the brane (large values of the
radial coordinate) the universe enters a slow inflationary phase,
in which the energy density is proportional to an inverse power of
the logarithm of the scale factor. As the scale factor grows the
induced energy density takes a constant value and the universe
enters a normal exponential expansion. For small values of the
scale factor the induced energy density scales as the inverse
powers of the scale factor and then the logarithmic terms take
over and the universe enters a slow exponential expansion.

The energy densities we calculated break down for some specific
values of the scale factor. This is a reflection of the fact that
the approximate solutions in the IR cannot be continued to the UV.
To answer the question if there is a true phase of "logarithmic
inflation" in which the universe inflates but with a slow rate, we
must resolve the problem of singularities, where our theory breaks
down. We are studying the problem numerically trying to solve the
equations of motion numerically \cite{Magg} and see if we can
smoth out the singularities. Then we can apply our technics for
calculating the effective energy density.

\section*{Acknowledgement}
We would like to thank A. Kehagias, E. Kiritsis and C. Bachas for
valuable discussions. Work partially supported by the NTUA program
Archimedes.

\section*{Note Added}
While this work was written up to its final form, the reference
\cite{Korean} appeared where a similar problem was studied, and it
was found that tachyonic background is less divergent than the one
without tachyon.

 \end{document}